\pdfoutput=1

\documentclass[pra,10pt,showpacs,superscriptaddress,footnoteinbib]{revtex4}
\usepackage{amsmath}
\usepackage{latexsym}
\usepackage{amssymb}
\usepackage{graphics,epstopdf}
\usepackage[colorlinks=true, citecolor=blue, urlcolor=blue ]{hyperref}
\usepackage{epsf,graphics,graphicx}

\newcommand{\ed}{\end{document}}
\newcommand{\beq}{\begin{equation}}
\newcommand{\eeq}{\end{equation}}

\begin{document}

\title{ Inertial spin Hall effect in noncommutative space}
\author{B. Basu\footnote{Electronic
address: {sribbasu@gmail.com}}${}^{}$, Debashree Chowdhury\footnote{Electronic
address:{debashreephys@gmail.com}}${}^{}$ and Subir Ghosh\footnote{Electronic
address: {sghosh@isical.ac.in}} ${}^{}$}
\affiliation{Physics and
Applied Mathematics Unit, Indian Statistical Institute,\\
 203
B.T.Road, Kolkata 700 108, India}


\begin{abstract}
In the present paper the study of inertial spin current(that  appears in an accelerated
frame of reference) is extended to Non-Commutative (NC) space. In the Hamiltonian framework, the Dirac Hamiltonian in an accelerating frame is
computed in the low energy regime by exploiting the Foldy-Wouthuysen scheme. The NC $\theta$-effect appears from
the replacement of normal products and commutators by Moyal $*$-products and $*$-commutators. In particular, the
commutator between the external magnetic vector potential and the potential induced by acceleration becomes
non-trivial. Expressions for $\theta$-corrected  inertial spin current and conductivity are derived explicitly. We have provided yet another way of experimentally measuring $\theta$. The $\theta$ bound is obtained from the out of plane spin polarization, which is experimentally observable.

\end{abstract}


\maketitle
keywords: {Spin Hall effect; Non inertial reference frame; Non commutative quantum mechanics; Spin polarization; Bound for non-commutative parameter.}

\section{Introduction:}
Systems living in Non-Commutative (NC) spacetime have created a lot of interest
in recent years
(for extensive reviews see e.g. \cite{d}). Although the
study of noncommutative geometry has a long history \cite{j}, it was revived in
a different perspective
after the seminal work of Seiberg and Witten \cite{sw} that showed its relevance
in certain low energy
limits of  String Theory. This work also paved the way for a systematics
framework where quantum field theories in
classical (or commutative) spacetime can be extended to NC spacetimes via the
Seiberg-Witten map \cite{sw}. This map
expresses a quantum field in NC spacetime in terms of a quantum field in
commutative spacetime in the form of a
power series in the NC parameter. In a relativistic framework, the spacetime
noncommutativity would be represented by
$ [x_{\mu}, x_{\nu}] = i\theta_{\mu\nu}$. For arbitrary $\theta_{\mu\nu}$,
higher (than quadratic) order time derivatives
will be generated through the Seiberg-Witten map, leading to a possible
violation of unitarity. To avoid this type of pathology it is  convenient to restrict the
noncommutativity to the spatial sector
only (i.e. $\theta _{0\mu}=0$), with the following phase space algebra,
\begin{equation}\label{nc}
[x_{i}, x_{j}] = i\theta_{ij}, ~
[x_{i}, p_{j}] = i\hbar\delta_{ij}, ~ [p_{i}, p_{j}] = 0.
\end{equation}
$\theta_{ij}$  an anti-symmetric real tensor, is the NC parameter. It should be mentioned that for (spacetime) constant
$\theta_{\mu\nu}$ Lorentz invariance is violated. An 
interesting alternative framework has been provided in \cite{carone}, (with
recent applications in \cite{carone1}), where $\theta_{\mu\nu}$ is elevated to a
Lorentz
tensor so that Lorentz invariance of the NC algebra is maintained although the
possibility of loss of unitarity from higher 
time derivatives through Seiberg-Witten map will remain. We have used the
simpler NC model of constant $\theta_{ij}$ (\ref{nc}) primarily because we
finally consider low energy non-relativistic
scenario. Operationally
effect of the Seiberg-Witten map on a quantum field theory is encoded
simply by replacing products of quantum field at the same spacetime point by
Moyal (or $*$) product as defined below,
\begin{eqnarray}
f(\vec{x})*g(\vec{x}) =
exp[\frac{i\theta_{ij}}{2}\partial_{xi}\partial_{yi}]f(\vec{x})g(\vec{y})|_{x=y}
.\label{moy}
\end{eqnarray}
In the above  $ f(\vec{x})$ and $ g(\vec{x})$ are two generic fields in
commutative space. Clearly the NC effects
generated from the $*$-product appear as higher derivative correction terms and
in the limit $\theta_{ij}\rightarrow 0$
the NC theory smoothly reduces to the original theory in commutative space.
Indeed the algebra (\ref{nc}) is
consistent with (\ref{moy}) where commutators are replaced by $*$-commutators,
\begin{equation}
 [A,B]_*=A*B-B*A
\label{*com}
\end{equation}

In low-energy limit, by considering the
one-particle sector of field theory on NC space one arrives at NC
extension of quantum mechanics.
To match with present experimental bounds, NC effects are expected to be very
small. Even then it is
interesting to study various effects of NC space on conventional systems. (For
different examples we cite
some of the works in \cite{l}.)

In this paper, we wish to discuss the effect of NC space on  inertial spin
current and spin polarization. The study of spin current is of recent interest
from different
perspectives \cite{50,51,52,54,56,58,60}. 
It were Dyakonov and  Perel
\cite{spinh}, who first predicted the spin Hall effect (SHE) and the effect was
theoretically  developed in \cite{sh1}. This effect is observed experimentally
in semiconductors and
metals
 \cite{sh2}. It is a form of anomalous Hall effect induced by spin.
Here a beam of particles separates in to up and down spin projections in the
presence of perpendicular electric field in analogy to Hall effect
where charges are separated in a beam passing through a perpendicular magnetic
field. SHE, occurs due to the spin orbit coupling(SOC) of electron, which is the
relativistic coupling of electron spin with orbital motion, 
 become an
active area of research\cite{50,51,52,54,56,58,60}. However, the studies
on the inertial effect of electrons has a long history
\cite{barnett,ein,tol,o} but the contribution of the spin-orbit coupling
 in accelerating frames has not much been addressed in the literature \cite{cb,bc}.
In the literature we find that there has been an elegant attempt to extend the
theory of spin current
in the inertial frame \cite{matsuoprl}. A theory has been  proposed describing
the direct coupling of the mechanical rotation and spin current and predicting
the spin current generation arising from rotational
motion.\cite{matsuoprl,c}. Recently, Spin Hall effect in NC space has been
studied in
\cite{db,b,dayi,jellal}.

Let us put the present work in its proper perspective. Apart from extending
previous results of inertial effects
of spin current and conductivity in NC space,(that is indeed of interest), the
main motivation was to provide a bound for 
$\theta$-the NC parameter involving an experimentally measurable parameter.
There have been many attempts to find measurable spatial NC effects. For example, a possibility of testing non-commutativity effect via cold Rydberg atoms \cite{ryd} was suggested. A lower bound \cite{bound} $1/\sqrt{\theta} \geq 10^{-6} GeV$ for the space NC parameter is obtained through the Aharonov-Bohm phase. The Aharonov -Casher phase has been utilized to find a limit for space NC parameter as $1/\sqrt{\theta} \approx 10^{-7}$\cite{bound1}. Besides, interesting results can be obtained through the discussion on NC quantum Hall effect using semi-classical constrained Hamiltonian \cite{dayi1}. Data on the spin Hall measurement \cite{SO} impose a different bound  on  $1/\sqrt{\theta}, $ than the one imposed by quantum Hall effect \cite{micu}.  
Indeed, we have succeeded in relating $\theta$ to the out of plane spin
polarization($s_{z}$), which is a measurable quantity.
In order to address the above-mentioned issues, in the
present work,  it is quite interesting to explore the conditions of the
spin-dependent
inertial forces and induced spin currents that appear in non-inertial
accelerating frames \cite{c} placed in the NC space.

The paper is organized as follows. Section II deals with a linearly
accelerating
frame in NC space.
The formalism adopted by us is explained in detail. Here we have used the Foldy
Wouthuysen transformation \cite{m} to get the non relativistic limit of Dirac
equation. We have dealt with the idea \cite{c} of interpreting the effect of
linear acceleration on an electron as induced effective electric field in this
section. The spin-orbit coupling resulting from this induced electric field
along with electric field due to the spin-orbit coupling generated from the
external electromagnetic field and noncommutativity produces the spin current
in our system. Here we follow the physically intuitive
approach of Chudnovsky \cite{n}, based on an extension of the Drude model
and  derive the  spin Hall current and conductivity in a NC
framework.
In section III we explain in detail how the noncommutativity affects the Rashba
\cite{rashba} like coupling parameter and the spin polarization. Lastly in
section IV we propose a new bound for the NC parameter $\theta$
from the expression of out of plane spin polarization. The paper ends with
Conclusions in section V.

\section{Inertial spin orbit coupling in presence of non commutativity}

Our general framework is the following: we first construct the Dirac equation in
a non-inertial frame,
following the work of Hehl and Ni \cite{o}. Subsequently we introduce the NC
effects. The essential idea in \cite{o}
is to introduce a system of orthonormal tetrad carried by the accelerating
observer. This in turn induces a
non-trivial metric and subsequently one rewrites the Dirac equation in the
observer's local frame where
normal derivatives are replaced by covariant derivatives derived from the
induced metric. A series of Foldy-Wouthuysen  Transformations  (FWT) \cite{m}
yield a non-relativistic approximation of
the Dirac Hamiltonian. Finally NC effects are incorporated by extending the
commutators appearing in FWT to $*$-commutators
defined in (\ref{*com}).

The  Dirac Hamiltonian in an arbitrary non-inertial frame with linear
acceleration and rotation  is given by \cite{o}
\begin{equation}
H = \beta mc^{2} +  c\left(\vec{\alpha}.(\vec{p}-\frac{e\vec{A}}{c})\right)
+\frac{1}{2c}\left[(\vec{a}.\vec{r})((\vec{p}-\frac{e\vec{A}}{c}).\vec{\alpha})
+
((\vec{p}-\frac{e\vec{A}}{c}).\vec{\alpha})(\vec{a}.\vec{r})\right] \\$$$$
+\beta m
(\vec{a}.\vec{r}) + eV(\vec{r}) - \vec{\Omega}.(\vec{L} + \vec{S}) \label{q} 
\end{equation}
where $\vec{a}$ and  $\vec{\Omega}$ are respectively the linear acceleration and
rotation frequency of the observer with respect to an inertial frame. $\vec{L}$
 and $\vec S$ are respectively  the angular momentum
($\vec{L} = \vec{r}\times \vec{p}$) and spin  of the Dirac particle.    $
\vec{A} $ denotes the
vector potential. The Dirac matrices $ \beta$,  $\alpha $ and the  spin
operator $\Sigma$ for 4-spinor are respectively given by
\begin{equation}\label{alfa}
\beta = \left( \begin{array}{cc}
I & 0 \\
  0 &-I
\end{array} \right),~~~~
\alpha= \left( \begin{array}{cc}
 0 & \sigma \\
  \sigma & 0
\end{array} \right), ~~~
 \Sigma = \frac{\hbar}{2}
 \begin{pmatrix}
  \sigma & 0\\
   0 & \sigma
 \end{pmatrix}
\end{equation}
In this section, we
drop the rotation term   $ \vec{\Omega}.(\vec{L}+ \vec{S}) $ to study only the effects of linear acceleration. For further
calculations one has to apply FWT \cite{o,m} on the Hamiltonian  in (4).

Let us outline the Foldy-Wouthuysen transformation (FWT) for NC
space in the present case.
The Dirac wave function is a four component spinor with the up and
down spin electron and hole components. Generically the energy gap between the
electron and hole is much larger than the energy scales associated with
condensed matter systems. Hence it is natural to  take the non relativistic
limit of Dirac equation. One
can achieve this by block diagonalization method of the Dirac Hamiltonian
exploiting FWT \cite{m}. For $
\vec{\Omega} = 0 $ in (4)
the Hamiltonian can be divided into  block diagonal and off
diagonal
parts denoted by $ \epsilon $ and $ O $ respectively. Thus the Hamiltonian can
be written as
\begin{eqnarray}
H &=& \beta mc^{2} + O +\epsilon ,\nonumber\\
 O &=& c\left(\vec{\alpha}.(\vec{p}-\frac{e\vec{A}}{c})\right)
+\frac{1}{2c}\left[(\vec{a}.\vec{r})((\vec{p}-\frac{e\vec{A}}{c}).\vec{\alpha})
+
((\vec{p}-\frac{e\vec{A}}{c}).\vec{\alpha})(\vec{a}.\vec{r})\right],\nonumber\\
\epsilon &=& \beta m (\vec{a}.\vec{r}) + eV(\vec{r}).
\label{fwh}
\end{eqnarray}
where $m$ is the mass of the Dirac particle and $\beta=\gamma_0$,
 $\alpha_i=\gamma_0\gamma_i$ are the Dirac matrices. Applying FWT on $H$ yields,
\begin{equation}
H_{FW} = \beta \left(mc^{2}+\frac{O^{2}}{2mc^{2}}\right)+ \epsilon
-\frac{1}{8m^{2}c^{4}}\left[O ,[O,\epsilon]\right].
\label{hfw1}
\end{equation}
Now comes the novel part of our work. As we have explained in the Introduction,
NC space effects can
be incorporated simply by replacing the (space dependent) products
and   brackets in the Hamiltonian (\ref{hfw1}) by $*$-products and $*$-brackets
(in particular
$O^2\rightarrow O*O,
[O,\epsilon ]\rightarrow [O,\epsilon ]_*$. We restrict ourselves to $ O(\theta
)$
results and find,
\begin{eqnarray}
O^{2} &=& c^{2}\frac{\left((\vec{p}-\frac{e\vec{A}}{c}) *
(\vec{p}-\frac{e\vec{A}}{c})\right)}{2m} -ce\hbar\vec{\Sigma} .\vec{B} +
ie^{2}\vec{\Sigma}.(\vec{A}\times_{*} \vec{A})\nonumber\\~
[O,\epsilon]_{*} &=& -ice\hbar \vec{\alpha} . \vec{\nabla} V(\vec{r}) - e^{2 }
\vec{\alpha} . [\vec{A} ,V(\vec{r})]_{*} -\beta
mic\hbar(\vec{\alpha}.\vec{a}) - \beta
me\vec{\alpha}.[\vec{A},\vec{a}.\vec{r}]_{*}\\~
\left[O ,[O,\epsilon]_{*}\right]_{*} &=& ec^{2}\hbar^{2} (\vec{\nabla}.\vec{E})
+iec^{2}\hbar^{2}\vec{\Sigma}.(\vec{\nabla}\times \vec{E})+
 2ec^{2}\hbar\vec{\Sigma}.(\vec{E}\times \vec{p})-
2ie^{2}c\vec{\Sigma}.([\vec{A},V(\vec{r})]_{*}\times \vec{p})\nonumber\\~
&-&\beta mc^{2}\hbar^{2}(\vec{\nabla}.\vec{a})-i\beta
mc^{2}\hbar^{2}\vec{\Sigma}.(\vec{\nabla}\times \vec{a}) - 2\beta
mc^{2}\hbar\vec{\Sigma}.(\vec{a}\times\vec{p}) + 2i\beta
cme\Sigma.\left([\vec{A},\vec{a}.\vec{r}]_{*}\times\vec{p}\right)
\end{eqnarray}
Terms like $ [ \vec{A},[\vec{A}, V(\vec{r})]_{*}]_{*}$, $ \vec{\nabla}.[\vec{A},
V(\vec{r})]_{*}$, $ \vec{\nabla}.[\vec{A},\vec{a}.\vec{r}]_{*}$etc are  $
O(\theta ^{2}) $
and hence dropped. The redshift effect of kinetic energy (\cite{o}) is also
neglected in the above calculations.
Adding all these terms, the FW transformed Hamiltonian on the NC
space takes the form,
\begin{equation}
H_{FW*} = \beta\left( mc^{2} + \frac{(p-\frac{e\vec{A}}{c}) *
(p-\frac{e\vec{A}}{c})}{2m}\right) + eV(\vec{r}) + \beta m (\vec{a}.\vec{r})$$$$
-\frac{e\hbar}{2mc}\vec{\Sigma}.\vec{B}-  \frac{e\hbar^{2}}{8m^{2}c^{2}}
(\vec{\nabla}.\vec{E}) -
\frac{ie\hbar^{2}}{8m^{2}c^{2}}\vec{\Sigma}.(\vec{\nabla}\times \vec{E})-
\frac{e\hbar}{4m^{2}c^{2}}\vec{\Sigma}.(\vec{E}\times \vec{p})$$$$
+\frac{\beta\hbar^{2}}{8mc^{2}}(\vec{\nabla}.\vec{a})
+ \frac{i\beta\hbar^{2}}{8mc^{2}}\vec{\Sigma}.(\vec{\nabla}\times \vec{a}) +
\frac{\beta\hbar}{4mc^{2}}\vec{\Sigma}.(\vec{a}\times \vec{p}) -\frac{\beta
e\hbar}{2mc}\vec{\Sigma}.\vec{B}_{\theta} +
\frac{e\hbar}{4m^{2}c^{2}}\vec{\Sigma}.(\vec{E}_{\theta}\times
\vec{p}) -\frac{i\beta
e}{4mc^{3}}\vec{\Sigma}.\left([\vec{A},\vec{a}.\vec{r}]_{*}\times\vec{p}
\right)\label{w}
\end{equation}
Let us simplify the Hamiltonian a little bit. As we are dealing with the
constant acceleration we can drop the terms $ (\vec{\nabla}.\vec{a})$ and
$\vec{\Sigma}.(\vec{\nabla}\times \vec{a})$.
 Consideration of constant electric field can help us leaving  the terms with
$(\vec{\nabla}\times \vec{E})$ and
$(\vec{\nabla}.\vec{E})$. Finally, we land up with the Hamiltonian as
\begin{equation}\label{hfw}
H_{FW*} = \left(mc^{2} + \frac{\left((p-\frac{e\vec{A}}{c}) *
(p-\frac{e\vec{A}}{c})\right)}{2m}\right) +
eV(\vec{r})-\frac{e\hbar}{2mc}\vec{\sigma}.\vec{B}+  m (\vec{a}.\vec{r}) $$$$
-\frac{e\hbar}{4m^{2}c^{2}}\vec{\sigma}.(\vec{E}\times \vec{p})+
\frac{\hbar}{4mc^{2}}\vec{\sigma}.(\vec{a}\times \vec{p})$$$$
 + \frac{e\hbar}{4m^{2}c^{2}}\vec{\sigma}.(\vec{E}_{\theta}\times
\vec{p}) - \frac{e\hbar}{4m^{2}c^{2}}\sigma.(\vec{E}_{\vec{a},\theta}\times
\vec{p}) -\frac{
e\hbar}{2mc}\vec{\sigma}.\vec{B}_{\theta}.
\end{equation}
where  $ \vec{E}_{\theta} = -\frac{ie}{c\hbar}[\vec{A},V]_{*},$ and ~~
$\vec{B}_{\theta} = -\frac{ie}{c\hbar}(\vec{A}\times_{*}\vec{A}) $ and $
\vec{E}_{\vec{a},\theta} = -\frac{ie}{c\hbar}[\vec{A},V_{a}(\vec{r})]_{*}$. Here
we consider a potential $V_{\vec{a}} = -\frac{m}{e}\vec{a}.\vec{r}.$  We make a
point here that the inertial effect of the linear acceleration on electron can
be interpreted as an  induced $effective$ $electric$ $field$ $\vec{E}_{\vec{a}}$
such that \beq \vec{E}_{\vec{a}} = \frac{m}{e}\vec{a},\eeq(where the induced
electric field $\vec{E}_{\vec{a}}$ is the gradient of some potential $
V_{\vec{a}}$). Introduction of this effective electric field $\vec{E}_{\vec{a}}$
generates an inertial spin-orbit term and its NC correction, apart
from the
spin-orbit term arising due to the external electric field (fifth term in the
right
hand side of (11). The above FW
Hamiltonian  on the NC space gives  the dynamics of an electron (or
hole with proper sign of
$e$) in the positive energy part of the full energy spectrum. Here
$\vec{\sigma}$ is the Pauli spin matrix.
To study  the dynamics
of the charged particles, the vector potential  ${\vec{A}}$ plays an important
role,
which  can be understood via the Aharonov-Bohm effect \cite{ahara}.
The choice of ${\vec A}$, the vector potential, will impose the condition on
${\vec B}$ and $\vec{B}_{\theta}={\vec A}\times_* {\vec A}.$ The choice put in
our analysis will be stated later.
\begin{equation}\label{hfw2}
H_{FW*}= \left(mc^{2} + \frac{\left((p-\frac{e\vec{A}}{c}) *
(p-\frac{e\vec{A}}{c})\right)}{2m}\right) +
eV(\vec{r})-\frac{e\hbar}{2mc}\vec{\sigma}.\vec{B} -  e V_{\vec{a}}(\vec{r})
$$$$
-\frac{e\hbar}{4m^{2}c^{2}}\vec{\sigma}.(\vec{E}\times \vec{p})+
\frac{e\hbar}{4m^{2}c^{2}}\vec{\sigma}.(E_{\vec{a}}\times \vec{p})$$$$
  +\frac{e\hbar}{4m^{2}c^{2}}\vec{\sigma}.(\vec{E}_{\theta}\times
\vec{p}) -
\frac{e\hbar}{4m^{2}c^{2}}\vec{\sigma}.(\vec{E}_{\vec{a},\theta}\times \vec{p})
-\frac{
e\hbar}{2mc}\vec{\sigma}.\vec{B}_{\theta}.
\end{equation}
We are now in a position to explain the underlying physics of the individual
terms of the right hand side of the Hamiltonian (13). The first two terms
describe the relativistic mass increase,  whereas the third  term is the
electrostatic energy and the fourth term is a magnetic dipole energy
which induces Zeeman effect. The fifth term arises due to linear acceleration in
the system.  The terms
 $\vec{\sigma}.(\vec{E}\times \vec{p}) ,$ $
\vec{\sigma}.(\vec{E}_{\vec{a}}\times \vec{p}) $ and
$\vec{\sigma}.(\vec{E}_{\theta}\times \vec{p}),$
$\vec{\sigma}.(\vec{E}_{\vec{a},\theta}\times \vec{p})$ are respectively
 the spin-orbit interaction terms and its correction due to noncommutativity.

To derive the equations of motion of the electron we follow the physically
intuitive approach of Chudnovsky \cite{n}, based on an extension of the Drude
model and its NC extension \cite{b}.
Collecting the dynamical terms and the terms due to spin orbit coupling, the
final Hamiltonian for the positive energy solution of spin $ \frac{1}{2}$
electron can now be read  as
\begin{equation}
H_{FW*} =  \frac{p^{2}}{2m} + eV(\vec{r}) - e V_{\vec{a}}(\vec{r})
- \frac{e\hbar}{4m^{2}c^{2}}\vec{\sigma}.(\vec{E}\times \vec{p}) +$$$$
\frac{e\hbar}{4m^{2}c^{2}}\vec{\sigma}.(\vec{E}_{\vec{a}}\times \vec{p}) +
\frac{e\hbar}{4m^{2}c^{2}}\vec{\sigma}.(\vec{E}_{\theta}\times \vec{p})
-\frac{e\hbar}{4m^{2}c^{2}}.\vec{\sigma}.(\vec{E}_{\vec{a},\theta}\times
\vec{p}) \label{1234}
\end{equation}
As we can see, in (14) we have neglected the
rest energy term.
Let us remind here that the space dependent potential $V(\vec{r})$  is the sum
of the external electric potential $ V_{0}(\vec{r})$ and the lattice electric
potential $ V_{l}(\vec{r}) $. On the non commutative space the Heisenberg's
equations of motion  can be obtained in the standard method as \cite{n}
\begin{eqnarray}
\vec{\dot{r}} & =& \frac{1}{i\hbar}[\vec{r}, H_{FW*}],
\\\vec{ \dot{p}} &=& \frac{1}{i\hbar}[\vec{p} , H_{FW*}].\label{08}
\end{eqnarray}
Consequently,  we have
\begin{eqnarray}
\vec{\dot{r}} &=& \frac{\vec{ p}}{m} +
\frac{e\hbar}{4m^{2}c^{2}}\left(\vec{\sigma}\times \vec{\nabla}
V(\vec{r})\right) -
\frac{ie^{2}}{4m^{2}c^{3}}\left(\vec{\sigma}\times [\vec{A}
,V(\vec{r})]_{*}\right)
-\frac{e\hbar}{4m^2c^{2}}(\vec{\sigma}\times \vec{\nabla} V_{\vec{a}}) +
\frac{ie^{2}}{4m^2c^3}\left(\sigma\times[\vec{A},V_{\vec{a}}(\vec{r})]_{*}
\right)\label{m} \nonumber \\
\vec{\dot{p}} &=& -e\vec{\nabla} V(\vec{r}) + e\vec{\nabla} V_{\vec{a}}(\vec{r})
-
\frac{e\hbar}{4m^{2}c^{2}}\vec{\nabla}((\vec{\sigma}\times
\vec{\nabla}V(\vec{r})).\vec{p}) \nonumber \\ & & + \frac{ ie^{2}
}{4m^{2}c^{3}}\vec{\nabla}
\left((\vec{\sigma}\times [\vec{A}, V(\vec{r})]_{*}).\vec{p}\right) +
\frac{e\hbar}{4m^{2}c^{2}}\vec{\nabla}\left((\vec{\sigma}\times \vec{\nabla}
V_{\vec{a}}(\vec{r})).\vec{p}\right) -
\frac{ie^{2}}{4m^{2}c^{3}}\vec{\nabla}\left((\vec{\sigma}\times[\vec{A},
V_{\vec{a}}(\vec{r})]_{*}).\vec{p}\right)\label{pqr}
\end{eqnarray}
From (\ref{m}) we can write
\beq  \vec{ p} =  m\vec{\dot{r}} - \frac{e\hbar}{4mc^{2}}(\vec{\sigma}\times
\vec{\nabla} V(\vec{r})) + \frac{ie^{2}}{4mc^{3}}(\vec{\sigma}\times
[\vec{A} ,V(r)]_{*}) + \frac{e\hbar}{4mc^{2}}\left(\vec{\sigma}\times
\vec{\nabla} V_{\vec{a}}(\vec{r})\right) -
\frac{ie^{2}}{4mc^3}\left(\sigma\times[\vec{A},V_{\vec{a}}(\vec{r})]_{*}
\right)\label{n1}  \eeq
The time derivative of (\ref{n1}) gives
\beq \vec{\dot{ p}} =  m\vec{\ddot{r}} -
\frac{e\hbar}{4mc^{2}}(\dot{\vec{r}} .\vec{\nabla})(\vec{\sigma}\times
\vec{\nabla} V(\vec{r})) + \frac{ie^{2}}{4mc^{3}}(\dot{\vec{r}}
.\vec{\nabla})\left(\vec{\sigma}\times [\vec{A} ,V(r)]_{*}\right) +
\frac{e\hbar}{4mc^{2}}(\dot{\vec{r}} .\vec{\nabla})\left(\vec{\sigma}\times
\vec{\nabla} V_{\vec{a}}(\vec{r}\right) - \frac{ie^{2}}{4mc^{3}}(\dot{\vec{r}}
.\vec{\nabla})\left(\vec{\sigma}\times[\vec{A},V_{\vec{a}}(\vec{r})]_{*} \right)
\eeq
Finally, the equation of motion has the form
\begin{eqnarray}
m\ddot{\vec{r}}=-e\vec{\nabla}V(\vec{r}) + e\vec{\nabla} V_{\vec{a}}(\vec{r})
- \frac{e\hbar}{4mc^{2}}\dot{
\vec{r}}\times\vec{\nabla}\times\left(\vec{\sigma}\times \vec{\nabla}
V(\vec{r})\right) +
\frac{ie^{2}}{4mc^{3}}\dot{ \vec{r}}\times\vec{\nabla} \times(\vec{\sigma}\times
 [\vec{A}, V(\vec{r})]_{*})\nonumber
\\+\frac{e\hbar}{4mc^{2}}\dot{ \vec{r}}\times\vec{\nabla}\times\left
(\vec{\sigma}\times \vec{\nabla} V_{\vec{a}}(\vec{r})\right) -
\frac{ie^{2}}{4mc^3}\dot{ \vec{r}}\times\vec{\nabla}
\times\left(\vec{\sigma}\times
 [\vec{A}, V_{\vec{a}}(\vec{r})]_{*}\right) \label{18}
\end{eqnarray}
or,
\begin{eqnarray}\label{lor1}
m\ddot{\vec{r}}&=& -e\vec{\nabla}\left(V(\vec{r}) - V_{\vec{a}}(\vec{r})\right)
-\dot{\vec{r}}\times \vec{\nabla}\times\nonumber \\
&&\left(\frac{e\hbar}{4mc^{2}}(\vec{\sigma}\times \vec{\nabla} V(\vec{r}))
-\frac{ie^{2}}{4mc^{3}}
(\vec{\sigma}\times [\vec{A}, V(\vec{r})]_{*})
 -\frac{e\hbar}{4mc^{2}}\left(\vec{\sigma}\times \vec{\nabla}
V_{\vec{a}}(\vec{r})\right) + \frac{ie^{2}}{4mc^3}(\vec{\sigma}\times [\vec{A},
V_{a}(\vec{r})]_{*})\right)
\end{eqnarray}
It is worth mentioning here that this spin dependent effective Lorentz force
noted in eqn.(\ref{lor1}), is responsible for the transport of the electrons in
the system on the NC space, and hence responsible for the spin current.
As expected, when we put $\vec{a}=0$ in eqn. (\ref{18}), $i.e$ the system is not
accelerated, but put in the NC framework, our force equation is similar to that
obtained in \cite{b},

From the expression of $\dot{\vec{r}}$  in (\ref{m}) we can write the linear
velocity in linearly accelerating frame on non commutative space as
\beq
\displaystyle {\dot{\vec{r}} = \frac{\vec{p}}{m} +
\vec{v}_{\vec{\sigma},~{\vec{a}},~\theta}} \eeq where
\begin{eqnarray}\label{mnn}
\displaystyle \vec{v}_{\vec{\sigma},~\vec{a},~\theta} &=&
\frac{e\hbar}{4m^{2}c^{2}} \vec{\sigma}\times\left[\vec{\nabla}
V(\vec{r}) - \vec{\nabla}
V_{\vec{a}}(\vec{r}) -  \frac{ie}{c\hbar}[\vec{A} , V(\vec{r})]_{*} +
\frac{ie}{c\hbar}[\vec{A} , V_{a}(\vec{r})]_{*}\right] \nonumber \\
&=&
-\frac{e\hbar}{4m^{2}c^{2}} \vec{\sigma}\times \vec
{\cal{E}}_{\vec{a},~\theta}\end{eqnarray}
is the effective spin dependent velocity in the NC space
with \beq\vec{\cal{E}}_{\vec{a},~\theta} = -\left [\vec{\nabla} V(\vec{r}) -
\vec{\nabla}
V_{\vec{a}}(\vec{r})  - \frac{ie}{c\hbar}[\vec{A} ,
V(\vec{r})]_{*}  + \frac{ie}{c\hbar}[\vec{A} , V_{a}(\vec{r})]_{*}\right] \eeq
being the $total$ $effective$ $electric$ $field$ present in the system. One
should note here that the velocity term in (\ref{mnn}) is dependent of the
potential $\vec{\nabla} V(\vec{r})$ and $\vec{\nabla}V_{\vec{a}}(\vec{r}),$ and
also on their NC corrections.
Thus the inertial effect on linear acceleration along with the non-commutativity
 produces the anomalous velocity term which in turn may yield the spin Hall
effect for the particular choice of the vector potential ${\vec A}$.

Our chosen gauge   $\vec{A} = (-x_{2}, x_{1}, 0)$ gives, $[\vec{A},
V(\vec{r})]_{*} = i\theta \vec{\nabla} V(\vec{r}), [\vec{A}, V_{a}(\vec{r})]_{*}
= i\theta \vec{\nabla} V_{a}(\vec{r}) .$
Hence, we yield
\begin{eqnarray}
\displaystyle \vec{ \cal{E}}_{\vec{a},~\theta} = -\left[\vec{\nabla} V(\vec{r})
-
\vec{\nabla} V_{\vec{a}} + \frac{\theta e}{c\hbar}\vec{\nabla} V(\vec{r})  -
\frac{\theta e }{c\hbar}\vec{\nabla} V_{a}(\vec{r})\right].
\end{eqnarray}

The polarized spin current  due to the total effective electric field $
\displaystyle \vec{\varepsilon}_{\vec{a},~\theta} $ is  thus given by
\beq \vec{j}_{s} = e~n~Tr\sigma_{i}\vec{v}_{\vec{\sigma},~\vec{a}}.\eeq
Thus the $i^{th}$ component of the spin current on the NC space is
given by
\beq | j^{i}_{s}| = \frac{n~e^{2}\hbar}{2m^{2}c^{2}}(\vec{S}\times
\vec{\cal{E}}_{\vec{a},~\theta})^{i}\eeq
where $\vec{S}$ is the spin vector.
From this expression of current one can find out the spin current in different
directions. This result can be compared with the result given in \cite{c}. It
can be easily checked from this expression that when $\theta=0$, $i.e.$
NC correction is absent, our result is in agreement with the results
given in \cite{c}. On the other hand, it can also be verified that when the
system is not accelerated, but put in a NC framework our analysis is
consistent with that of \cite{b} with a different choice of vector potential.

The next job is to evaluate the explicit expression for the spin Hall
conductivity in an accelerated system.
With a careful observation of the force equation one can put further analysis on
different components of the total current produced in a accelerating system
moving in a NC space by adopting the averaging methodology followed in \cite{n}.
With our choice of gauge, one can
write the force equation  (\ref{lor1}) as
\begin{eqnarray}\label{lor}
m\ddot{\vec{r}} &=& \vec{F}_{0}+ \vec{F}_{\vec{\sigma}}\\
&=& \vec{F}_{0} + F_1(\vec{\sigma}) + F_2(\vec{\sigma},\theta)
\end{eqnarray}
where $ \vec{F}_{0}$ is the  spin independent part of the force and
$\vec{F}_{\vec{\sigma}}$ is the spin dependent part of the total spin force,
which actually corresponds to two parts, (i) spin dependent but $\theta$
independent part $\vec{F}_1(\vec{\sigma})$ and
(ii)$\vec{F}_2(\vec{\sigma},\theta)$ which is the noncommutative correction to
the spin dependent part.
The explicit forms of the above mentioned terms are
\begin{eqnarray}\label{sp0}
\vec{F}_0 & =& -e\vec{\nabla}V(\vec{r})+
e\vec{\nabla}V_{\vec{a}}(r)=-e\vec{\nabla}V_{tot},\\
\vec{F}_1(\vec{\sigma}) &=& -\frac{e\hbar}{4mc^{2}}\dot{ \vec{r}}\times
(\vec{\nabla}\times(\vec{\sigma}\times \vec{\nabla} V(\vec{r}))
+
\frac{e\hbar}{4mc^{2}}\dot{\vec{r}}\times(\vec{\nabla}\times(\vec{\sigma}\times
\vec{\nabla}V_{\vec{a}}))\label{s}
\\ \vec{ F}_2(\vec{\sigma},\theta) &=& - \frac{\theta e^{2}}{4mc^{3}}\dot{
\vec{r}}\times
(\vec{\nabla}\times(\vec{\sigma}\times\vec{\nabla} V(\vec{r}) ) + \frac{\theta
e^{2}}{4mc^{3}}\dot{ \vec{r}}\times
(\vec{\nabla}\times(\vec{\sigma}\times\vec{\nabla} V_{a}(\vec{r}) )
\end{eqnarray}

Here we have neglected $1/c^{4}$ terms. We can easily compare the form of the
force obtained in (\ref{lor1}), with the Lorentz force on the charge $e$. The
spin dependent part of this effective Lorentz force, $i.e$
$\vec{F}_1(\vec{\sigma})$ and $\vec{F}_2(\vec{\sigma}, \theta)$ are responsible
for the inertial spin current of the system on the NC space. One can write the
total Lorentz force as
\beq \vec{F}_0 + \vec{F}_{\vec{\sigma}} = e\vec{E}_{tot} +
\frac{e}{c}(\dot{\vec{r}} \times\vec{ B}(\vec{\sigma}))\label{sigma},\eeq where
the spin Lorentz force is
\beq  \vec{F}_{\vec{\sigma}} =  \frac{e}{c}(\dot{\vec{r}} \times
\vec{B}(\vec{\sigma}))\label{sigma1}.\eeq
This spin dependent Lorentz force is responsible for the spin current produced
in the system. $ \vec{B}(\vec{\sigma})$, the $effective$ $magnetic$ $field$
appearing in the spin space can be read as
\begin{eqnarray}\label{asigma}
\vec{B}(\vec{\sigma}) &=& \vec{\nabla}\times \vec{A}(\vec{\sigma})\\
&=& \vec{\nabla} \times (\vec{A}_{1}(\vec{\sigma}) + \vec{A}_{2}(\vec{\sigma} ,
\theta)),
\end{eqnarray}
where the forms of the vector potentials are explicitly given by
\begin{eqnarray}
\vec{A}_{1}(\vec{\sigma}) &=&   -\frac{\hbar}{4mc}(\vec{\sigma}\times
\vec{\nabla} V(\vec{r}))+ \frac{\hbar}{4mc}(\vec{\sigma}\times
\vec{\nabla}V_{\vec{a}}) \\
\vec{A}_{2}(\vec{\sigma} , \theta) &=& - \frac{\theta
e}{4mc^{2}}(\vec{\sigma}\times\vec{\nabla} V(\vec{r})) +
\frac{e\theta}{4mc^{2}}(\vec{\sigma}\times\vec{\nabla} V_{\vec{a}}(\vec{r}) ).
\end{eqnarray}
Finally, the Hamiltonian (14) can be written as
\beq H_{FW*} = \frac{1}{2m}(\vec{p} -\frac{e}{c}\vec{A}(\vec{\sigma}))^{2} +
eV_{tot} \eeq
Proceeding further to deduce the spin current, the averaging methodology
followed in \cite{n} is adopted.
From eqn. (\ref{lor1}), it is observed that the contribution from $\vec{ F}(
\vec{\sigma})$ in comparison to $ \vec{F}_{0}$ ~ is very small. One can treat
this  as a perturbation in the next part of our calculation \cite{n}. Breaking
into different parts, the solution of
equation (\ref{18}) can be written as $\dot{ \vec{r}} = \dot{ \vec{r}}_{0} +
\dot{ \vec{r}}_{\vec{\sigma}}, $ where $\dot{ \vec{r}}_{\vec{\sigma}}$ has two
parts, arising  from the effect of force $ {\vec{F}}_1(\vec{\sigma}) ,
\vec{F}_2(\vec{\sigma},\theta).$ So we can write
\beq  \dot{ \vec{r}}_{\vec{\sigma}}  =  \dot{ \vec{r}}_1(\vec{\sigma})  +  \dot{
\vec{r}}_{2}(\vec{\sigma}, \theta) \eeq
If the relaxation time $\tau$ is independent of $\vec{\sigma}$ and  for the
constant total electric field  $\vec{E}_{eff}$, following \cite{n} we can write,
\beq \langle\dot{ \vec{r}}_{0}\rangle =
-\frac{\tau}{m}\left\langle\frac{\partial V_{eff}}{\partial r}\right\rangle =
\frac{e\tau}{m}\vec{E}_{eff}, \label{r0dot} \eeq
 where we denote  $\vec{E}_{eff} = -e \vec{\nabla}\left(V_{0}(\vec{r}) -
V_{\vec{a}}(\vec{r})\right).$
\begin{equation}
~~~~~~~~~~~~~~~~~\left\langle\dot{\vec{r}}_{1}(\vec{\sigma})\right\rangle = -
\frac{\hbar
e^{2}\tau^{2}}{2m^{3}c^{2}}\vec{E}_{eff}\times\left\langle\frac{\partial
}{\partial r} \times (\vec{\sigma}
\times\frac{\partial V_{l}}{\partial r} )\right\rangle
+\frac{\hbar\tau^{2}e^{2}}{2m^{3}c^{2}}
\vec{E}_{eff}\times\left\langle\frac{\partial }{\partial r} \times
(\vec{\sigma} \times \frac{\partial V_{\vec{a}}}{\partial
r})\right\rangle\label{30}\\
\end{equation}
\begin{equation}\label{deb}
\left\langle\dot{\vec{r}}_2(\vec{\sigma},\theta)\right\rangle = -
\frac{\theta
e^{3}\tau^{2}}{4m^{3}c^{3}}\vec{E}_{eff}\times\left\langle\frac{\partial
}{\partial r} \times(\vec{\sigma}
\times  \frac{\partial V_{l}}{\partial r} )\right\rangle +
\frac{e^3\tau^2\theta}{4m^3c^3}\vec{E}_{eff}\times\left\langle\frac{\partial
}{\partial r} \times(\vec{\sigma}
\times  \frac{\partial V_{a}}{\partial r} )\right\rangle
\end{equation}
In the above expression of
$\left\langle\dot{\vec{r}}_{1}(\vec{\sigma})\right\rangle$ and
$\left\langle\dot{\vec{r}}_{2}(\vec{\sigma} , \theta)\right\rangle,$ a term is
present which represents the volume average of electrostatic potential
$\partial_{i}\partial_{j}V_{l}(r)$. In the study \cite{n} of the spin Hall
effect of $Al$, which is a
cubic lattice, the only contribution permitted by symmetry is
\beq \left\langle\frac{\partial^{2}V_{l}}{\partial r_{i}\partial
r_{j}}\right\rangle = \mu \delta_{ij},\label{sym}\eeq
where $\mu$ is a constant depending on the system. However, on NCS, within the
first order correction \cite{b}, we can use
(\ref{sym}) and the value of $[\vec{A}, V(\vec{r})]_*$ to obtain,
\beq \left\langle\dot{\vec{r}}_{1}(\vec{\sigma})\right\rangle = \frac{\hbar
e^{2}\tau^{2} \mu }{4m^{3}c^{2}}(\vec{\sigma}\times \vec{E}_{eff})
\label{r1dot}\eeq
and \begin{equation}\label{r2dot}
\left\langle\dot{\vec{r}}_2(\vec{\sigma},\theta)\right\rangle =
\frac{\theta e^{3}\tau^{2}\mu}{4m^{3}c^{3}} (\sigma\times\vec{E}_{eff})
\end{equation}
One may note that the second term in the right hand side of (\ref{30}) and
(\ref{deb}) vanishes for constant
acceleration and so contribution from that term is zero in (\ref{r1dot}) and
(\ref{r2dot}). It is interesting to notice that this term will contribute when
the system is under non linear acceleration.

To calculate the spin current we now introduce the spin polarization vector
$\vec{\lambda}=\langle \vec{\sigma}\rangle.$
The density  matrix of the charge carriers in spin space can be written as
\beq\rho~^{s} = \frac{1}{2}\rho(1 + \vec{\lambda}.\vec{\sigma})\label{deb1}\eeq
where $\rho$ is the total charge concentration.
Using eqn (\ref{deb1}) and the eqns(\ref{r0dot}), (\ref{r1dot}) and
(\ref{r2dot}) we derive the total spin current  as
\beq \vec{j} = e\left\langle\rho^{s}\vec{\dot{r}}\right\rangle = \vec{j}^{o
,\vec{a}} + \vec{j}^{s}(\vec{\sigma}) +
\vec{j}^{s}(\sigma ,\theta) \eeq
The various components of this current are given by
\beq \vec{j}^{o , \vec{a}} = \frac{e^{2}\tau \rho}{m}\vec{E}_{eff}, \label{jo}
\eeq
\beq \vec{j}^{s}(\vec{\sigma})  = \left(\frac{\hbar
e^{3}\tau^{2}\rho\mu}{2m^{3}c^{2}}\right)(\vec{\lambda}\times
\vec{E}_{eff}),\label{js} \eeq
\beq \vec{j}^{s}(\vec{\sigma} ,\theta) =
\left(\frac{e^{4}\tau^{2}\mu\rho\theta}{2m^{3}c^{3}}\right)(\vec{\lambda}\times
\vec{E}_{eff})\label{jthe} \eeq
of which is
$\vec{j}^{s}(\vec{\sigma})$ is $\theta$ independent and
$\vec{j}^{s}(\vec{\sigma} ,\theta)$ is $\theta$ dependent part of current.

The term $ \vec{j}^{o , \vec{a}}$ has two parts as  $\vec{j}^{o}$ and
$\vec{j}^{\vec{a}}$,where
\beq \vec{j}^{o } = \frac{e^{2}\tau \rho}{m}\vec{E} \eeq

\beq \vec{j}^{\vec{a}} =  \frac{e^{2}\tau \rho}{m}\vec{E}_{\vec{a}}\label{09}.
\eeq In the expression of (\ref{09}), we know that $\vec{E}_{\vec{a}} =
\frac{m\vec{a}}{e}.$ So when there is no acceleration in the system $
\vec{E}_{\vec{a}}$ becomes zero and there will be no spin current due to
acceleration. Inserting the value of $\vec{E}_{\vec{a}}$ in (\ref{09}) one can
get \beq \vec{j}^{\vec{a}} =  e\tau \rho\vec{a} = e\tau \rho a\hat{a},\label{05}
 \eeq where $\vec{a} = a\hat{a}$, $\hat{a}$ is the unit vector in the direction
of $\vec{E}_{\vec{a}}$. Thus one can define a parameter which is analogue of the
$conductivity$ as \beq \sigma^{\vec{a}}_{H} =  e\tau \rho a\eeq which appears as
an effect of the effective electric field generated due to the inertial effect
of linear acceleration in the system.
We can derive the corresponding spin Hall conductivities from the expressions of
the spin currents (\ref{jo}), (\ref{js}) and (\ref{jthe}) respectively as
\begin{eqnarray}\label{mdm}
\sigma_{H} &=& \frac{e^{2}\tau \rho}{m}\nonumber\\
\sigma^{s}_{H} &=& \frac{\hbar e^{3}\tau^{2}\rho\mu}{2m^{3}c^{2}}\nonumber\\
 \sigma^{s}_{H\theta} &=& \frac{e^{4}\tau^{2}\mu\rho\theta}{2m^{3}c^{3}}
 \end{eqnarray}
$\sigma^{s}_{H\theta}$ is the conductivity  arising due to the NC
correction. If we consider the effective electric field acts in the $ z $
direction, the component of spin current are as follows
\begin{eqnarray}\label{cond1}
\hat{j}^{s}_{x}(\vec{\sigma}) &=& (\sigma^{s}_{H} +
\sigma^{s}_{H\theta})E_{eff}\\
\hat{j}^{s}_{y}(\vec{\sigma}) &=& - (\sigma^{s}_{H} +
\sigma^{s}_{H\theta})E_{eff},
\end{eqnarray}
where $E_{eff}$ is the absolute value of $\vec{E}_{eff}.$ It is worthwhile to
point out here that our result differs from
a similar result of \cite{b} because in \cite{b} the authors took an explicit
form for $\vec A$ that can be
trivially gauged away.

The non commutative correction ratio $R$ can be written as
\beq R = \frac{\sigma^{s}_{H\theta}}{\sigma^{s}_{H}} = \frac{e\theta}{c\hbar}
\eeq
Now from (\ref{mdm}) we can evaluate the ratio of charge and spin hall
conductivities as
\beq\frac{\sigma^{s}_{H} + \sigma^{s}_{H\theta} }{\sigma_{H}} = \frac{\hbar
e\tau\mu}{2m^{2}c^{2}}(1+\frac{e\theta}{c\hbar} ) = \frac{\hbar
e\tau\mu}{2m^{2}c^{2}}(1+R) . \label{dib}\eeq
If we put $\theta $ equal to zero in (\ref{dib}), the result is similar as
\cite{n}. Our results can yield a bound for $\theta $ provided $R$ - the ratio
of conductivities are experimentally measurable.
\vspace{.5 cm}

\section{Spin polarization in NC space}

In this section, we consider only
the $\theta$ dependent terms, in the Dirac Hamiltonian (14) for a linearly
accelerating frame. 
\begin{equation}
H_{\theta} = \frac{\vec{p}^{2}}{2m} +
\frac{ie^2}{4m^{2}c^{3}}\vec{\sigma}.([\vec{A} ,
V_{a}(\vec{r})]_{*}\times\vec{p})
\end{equation}
In terms of the $induced$ $effective$ $field$ $\vec{E}_{\vec{a}, \theta}$ the
spin-orbit Hamiltonian for the NC part only is obtained as
 \begin{equation}\label{kspace}
H_{\theta} = \frac{\hbar^{2}\vec{k}^2}{2m} +
\frac{e\hbar^2}{4m^{2}c^{2}}\vec{\sigma}.(\vec{k}\times
\vec{E}_{\vec{a},\theta})
\end{equation}
where $\vec{E}_{a,\theta}$ represents the induced effective electric field
present in the system due to acceleration and noncommutativity and $\vec{p} =
\hbar \vec{k}$ is the crystal momentum , $m$ is the electrons mass.

The spin-orbit coupling due to combined action of noncommutativity and linear
acceleration, can now be analyzed by the same framework as the standard
Hamiltonian with SOC. The special relativity arguments can qualitatively explain
the effect of SOC.
For electrons moving through a lattice, the electric field $\vec{E}$  is Lorentz
transformed to an effective magnetic field $(\vec{k}\times \vec{E})\approx
\vec{B}(\vec{k})$ in the rest frame of the electron. Thus from (\ref{kspace}) we
can write
\beq H_{\theta} = \frac{\hbar^{2}\vec{k}^2}{2m} + \gamma
\vec{\sigma}.\vec{B}_{\vec{a},\theta}(\vec{k})\label{kha}.\eeq
Here, $\gamma$ is the coupling strength and  $ \vec{B}_{\vec{a},\theta}$ is the
effective magnetic field in the momentum space due to
$\vec{E}_{\vec{a},\theta}$.
For each value of $\vec{k}$, the spin degeneracy of electrons split into two
subbands $|\pm\rangle.$ 

In laboratory frame the magnetic field axis rotates as one moves from one point
to other.
A local transformation may rotate the frame in such a way that the spin
axis(along z axis) points along  the
unit vector $ \hat{n}_{a,\theta}(\vec{k}),$
where
\beq \hat {n}_{a,\theta}(\vec{k}) =
\frac{\vec{B}_{\vec{a},\theta}(\vec{k})}{|\vec{B}_{\vec{a},\theta}(\vec{k})|}
\label{na}.\eeq
 Here $hat$ above $n$ denotes that $\hat {n}_{a,\theta} $ is a unit vector.

For a particular choice  $ \vec{a}= (0, 0, a_{z}\hat{z})$, $
\vec{E}_{\vec{a},\theta} = (0,0,\frac{e\theta}{c\hbar}\vec{E}_{a,z}\hat{z}),$ we
can write  $\vec{B}_{a,\theta}(\vec{k}) = (\frac{e\theta}{c\hbar})(a_{z}k_{y},
-a_{z}k_{x}).$
From (\ref{kha}), the SOC Hamiltonian for this accelerating frame in
NC space can be written as
\begin{eqnarray} \label{rasba}
H_{\theta} &=& \frac{\hbar^{2}\vec{k}^2}{2m} -
\alpha\frac{e\theta}{c\hbar}(k_{x}\sigma_{y} - k_{y}\sigma_{x}) \\
&=& \frac{\hbar^{2}\vec{k}^2}{2m} - \alpha_{\theta}(k_{x}\sigma_{y} -
k_{y}\sigma_{x})
\end{eqnarray} where \beq\alpha _{\theta} = \alpha \frac{e\theta}{c\hbar}=
R\alpha .\eeq Here $R = \frac{e\theta}{c\hbar}$ and $\alpha$ contains the effect
of $a_{z}$ only  \cite{cb}, whereas $\alpha_{\theta}$ is the coupling strength of the
accelerating system on the NC
space.
The spin-orbit Hamiltonian (\ref{rasba}) is similar to the well known Rashba
Hamiltonian \cite{rashba} where
the coefficient $\alpha_{\theta}$ represents a Rashba like coupling in NC space
with the presence of  acceleration. However, in the usual Rashba coupling,
structural inversion asymmetry is responsible for the generation of internal
field, whereas in our formulation, responsibility lies with the electric field
induced  due to acceleration further modified by  NC effects. Due to the
presence of noncommutativity the strength of the coupling constant $\alpha$
modified to $\alpha_{\theta}$ in a accelerating system. Again as expected, it is
interesting to note that for $\theta = 0$ in the above equations we retrieve 
back the same result as in \cite{cb}. As the Rashba coupling strength have
enormous effects on condensed matter systems, we can view a lot of interesting
results with this Rashba like coupling strength in NC framework.

Dealing of the spin-orbit Hamiltonian in a NC space with a time
dependent acceleration $\vec{a}(t)$ provides us with some interesting results.
For a time dependent acceleration \cite{c} the induced effective time dependent
electric field, given by $\vec{E}_{\vec{a}}$,
is also time dependent.
Thus the time dependent Hamiltonian (\ref{kha})  is  \beq  H_{\theta}(t) =
\frac{\hbar^{2}\vec{k}^2}{2m}+ \gamma
\vec{\sigma}.\vec{B}_{a,\theta}(\vec{k}(t)).\label{time} \eeq

Finally, the expression for out of plane spin polarization in the NC space can
be derived. The acceleration of the carriers along with the time dependence of
the spin orbit Hamiltonian  generates an additional component
\cite{fujita,cb}  $\vec{B}_{\perp} = (\vec{\dot{n}}_{\vec{a},\theta}\times
\hat{n}_{\vec{a},\theta}),$
in addition to the effective magnetic field $\vec{B}_{a,\theta}(\vec{k}),$ which
basically explains the origin of the out of plane spin polarization. 
Thus following the methodology in \cite{fujita}, the out of plane spin
polarization in NC space is given by
\begin{eqnarray}\label{szt}
s^{\theta}_{z}\approx &\pm&
\frac{1}{|\vec{B}_{\vec{a},\theta}(\vec{k})|}\frac{\hbar}{2}(\vec{\dot{n}}_{\vec
{a},\theta}\times \hat{n}_{\vec{a},\theta}).\hat{z}\\
&=& \pm\frac{\hbar^{2}}{2\alpha_{\theta}
p}\frac{\hbar}{2}(\vec{\dot{n}}_{\vec{a},\theta}\times
\hat{n}_{\vec{a},\theta}).\hat{z}
\end{eqnarray}
The above expression shows a very important result of our calculation. It states
that if the electron is simultaneously in a accelerating frame and also in a non
commutative space, the out of plane spin polarization depends on a $\theta$
dependent Rashba like coupling parameter $\alpha_{\theta}$. We get back the
results of  \cite{cb}, when we substitute $\theta =0.$ One should notice here
that the expression of $s^{\theta}_{z}$ depends on the exact configuration of
the unit vector ${\hat{n}}_{\vec{a},\theta}$ and the time derivative of $\hat{n}_{\vec{a},\theta}$.

\section{Discussion and a new bound for $\theta$}
Using the expression of the out of plane spin polarization in this section we
want to propose a new bound for $\theta$. In the commutative sector the out of plane spin polarization vector is given by \cite{cb}
\begin{equation}\label{szm}
s_{z}\approx  \pm\frac{\hbar^{2}}{2\alpha
p}\frac{\hbar}{2}(\vec{\dot{n}}_{\vec{a}}\times
\hat{n}_{\vec{a}}).\hat{z}
\end{equation}
As the $\theta$ parameter is considered as a constant, the direction of
$\hat{n}_{a, \theta}$ in (\ref{na}) is same as $\hat{n}_{a}$ in the reference
\cite{cb}. 
We can then write the NC correction ratio by comparing the expressions of (\ref{szt}) and (\ref{szm}) as
\beq Q = \frac{s_{z}}{s^{\theta}_{z}} = \frac{\alpha_{\theta}}{\alpha} =
\frac{e\theta}{c\hbar}= R \label{thet}.\eeq  
A new bound for $\theta$ can be set by considering $Q$ to be of the order of one. Thus we can set a
lower  limit for $\theta $, using the relation (\ref{thet}) as
$\frac{1}{\sqrt{\theta}} \geq 10^{-12}$Gev. This is one of our major results.
\section{conclusion}
To conclude, a very general framework of a Dirac electron in external
electromagnetic field has been
considered that lives in a non-inertial as well as NC frame. The
non-inertial effects are introduced
in the Dirac equation and subsequently Foldy-Wouthuysen transformations are
exploited to reduce the
system to a
non-relativistic regime. Subsequently  non-commutativity effects are introduced by
replacing commutator brackets by $*$-commutator brackets
 (or Moyal brackets) and the dynamics of the resulting system is analyzed in
detail from a spin Hall effect
perspective. In particular, generalized expressions for the spin current and
subsequent spin Hall conductivities are computed to the lowest non-trivial order
in $\theta $ - the non-commutativity
parameter. The intriguing  part of our result is the fact that the acceleration
$\vec a$ and
non-commutative $\theta $ effects are explicitly entangled in the spin current
expressions (\ref{js},\ref{jthe}) however,
surprisingly, this is not manifested in spin Hall conductivities (\ref{cond1})
(as they are conventionally defined).
The non-inertial and non-commutative effects in spin Hall conductivities are
mutually exclusive, (at least to the order of approximation we
have considered). Here the experimentally relevant part of our analysis is that
we get a $\theta$ dependent Rashba like coupling parameter in the non
commutative space which causes a decrement of the out of plane spin
polarization. We have suggested a novel way of providing a bound for $\theta$ through $s^{\theta}_{z}$ that
can be subjected to  experiment. Our predicted numerical value for the $\theta$-bound agrees with existing results.

\end{document}